\documentclass[showkeys,superscriptaddress,notitlepage]{revtex4-1}
\usepackage{amsmath,amssymb,amsfonts,graphics,graphicx,dcolumn,bm}

\usepackage[breaklinks,colorlinks=true]{hyperref}
\usepackage{graphicx}
\usepackage[right]{lineno}
\usepackage{nameref,hyperref}
\usepackage{geometry}
\usepackage{epstopdf}

\begin{document}

\title{Sex differences in social focus across the lifecycle in humans}

\author{Kunal Bhattacharya}
\email[Corresponding author; ]{kunal.bhattacharya@aalto.fi}
\author{Asim Ghosh}
\author{Daniel Monsivais}
\affiliation {Department of Computer Science, Aalto University School of Science, P.O. Box 15400, FI-00076 AALTO, Finland}
\author{Robin I. M. Dunbar}
\affiliation {Department of Experimental Psychology, University of Oxford, South Parks Rd, Oxford, OX1 3UD, United Kingdom}
\affiliation {Department of Computer Science, Aalto University School of Science, P.O. Box 15400, FI-00076 AALTO, Finland}
\author{Kimmo Kaski}
\affiliation {Department of Computer Science, Aalto University School of Science, P.O. Box 15400, FI-00076 AALTO, Finland}
\affiliation {Department of Experimental Psychology, University of Oxford, South Parks Rd, Oxford, OX1 3UD, United Kingdom}

\date{\today}

\begin{abstract}
Age and gender are two important factors that play crucial roles in the way organisms allocate their social effort. In this study, we analyse a large mobile phone dataset to explore the way lifehistory influences human sociality and the  way social networks are structured. Our results indicate that these aspects of human behaviour are strongly related to the age and gender such that younger individuals have more contacts and, among them, males more than females. However, the rate of decrease in the number of contacts with age differs between males and females, such that there is a reversal in the number of contacts around the late 30s. We suggest that this pattern can be attributed to the difference in reproductive investments that are made by the two sexes. We analyse the inequality in social investment patterns and suggest that the age and gender-related differences that we find reflect the constraints imposed by reproduction in a context where time (a form of social capital) is limited. 
\end{abstract}


\maketitle


\section*{Introduction}
Most species, and humans in particular, exhibit striking changes in social style across the lifecycle, in most cases as a consequence of a shift in emphasis from development to reproduction. In humans, a greatly extended period of socialization, combined with a virtually unique period of post-reproductive (grandparental) investment, adds significant complexity to this. Although this much is obvious from casual observation, we actually know very little about the relative investment that individuals make as they age, or how this differs between the sexes. The last decade has seen a rapid growth and development in the information and communications technology (ICT), which has increasingly aided humans to connect to each other. Among the different channels that have become accessible, mobile phone communication is perhaps the most prominent as regards the number of users \cite{international2014world}. This is the reason 
mobile phone call data records (CDRs) have increasingly been used to study various aspects of human behaviour \cite{wang2011human,bagrow2011collective,blondel2015survey}. For example, from these CDRs one can construct egocentric networks that in turn allow one to undertake detailed studies of ego-alter relationships and the patterns of social investment that individuals make in the different members of their social networks \cite{saramaki2015seconds,miritello2013time,saramaki2014persistence}. 

\begin{figure}[t]
\begin{center}
\includegraphics[width=0.8\textwidth]{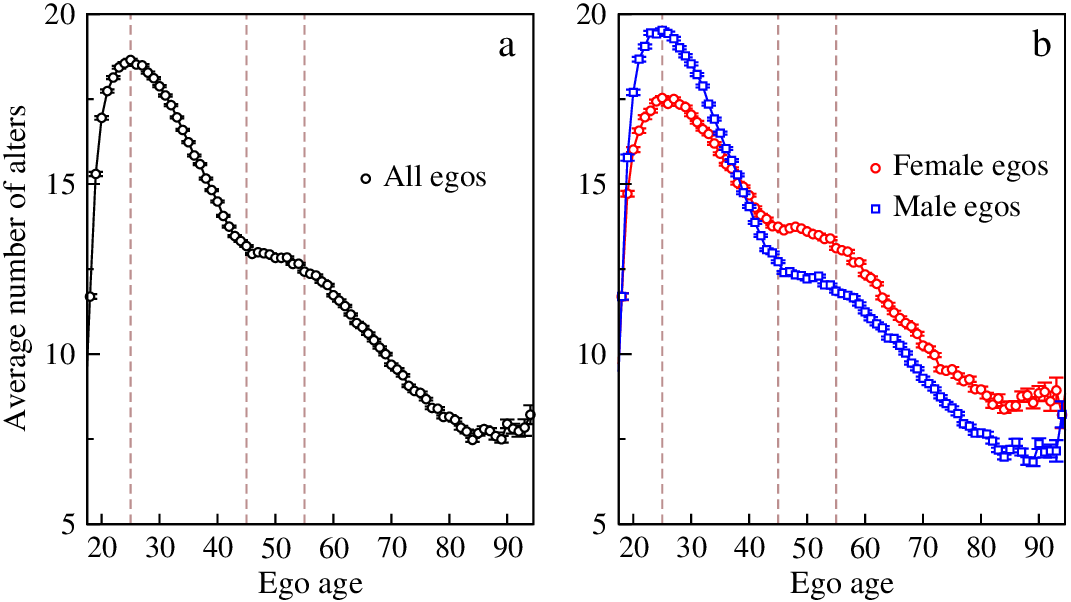}
\end{center}
\caption{The variation of the average number of alters with the age of the ego (in years). The measurement is done in the time window of a month and then averaging is carried over the 12 months. (a) Egos are considered irrespective of their sex. (b) The behaviour is shown for separate sexes. Male and female egos are denoted by blue squares and red circles, respectively. The error bars span the 95 percent confidence interval. The dashed lines in the background are used to demarcate different regimes of behaviour (discussed in Results).}
\label{fig-1}
\end{figure}

Previous studies have shown that individuals' telephone communication (landline and mobile) rates correlate with their face-to-face interactions \cite{roberts2011communication,saramaki2014persistence}. Both the age and gender of individuals have been found to be important factors influencing their communication patterns in mobile phone networks: the gender and age preferences of egos for their alters, for example, have been found to correlate with their geographic proximity \cite{jo2014spatial}. Furthermore, the dynamics of activation and deactivation of ties between individuals have been found to be different for the two genders and across age \cite{miritello2013limited}. In general, homophily and heterophily, which are known to be factors shaping human social interactions \cite{mcpherson2001birds,ip2006birds}, have turned out to be important in mobile phone communications, at least as regards to the gender preferences of an ego. In a previous study it was found that for younger egos, the most contacted alter is of the opposite sex \cite{palchykov2012sex}. Taken together, this suggests that, whatever their limitations might be, mobile phone data provide valid and reliable insights into human social patterns. 

In this paper, we analyze a large mobile phone dataset and study the structure of the individual level or egocentric networks. In general, we focus on their static structure for different non-overlapping periods, ranging from a month to a full year. In everyday life for both humans \cite{sutcliffe2012relationships} and other primates \cite{dunbar2014reproductive,dunbar2010bondedness}, time represents a direct measure of relationship quality. And, because time is limited and social investment is costly in terms of time \cite{lehmann2007group,dunbar2009time}, individuals are forced to choose how to distribute that time across the members of their network \cite{sutcliffe2012relationships,miritello2013limited,saramaki2014persistence}. Here, we use cross-sectional data on the frequency and duration of phone calls to examine how the pattern of social investment varies across the lifecycle in the two sexes.

\begin{figure}[h]
\begin{center}
\includegraphics[width=0.9\textwidth]{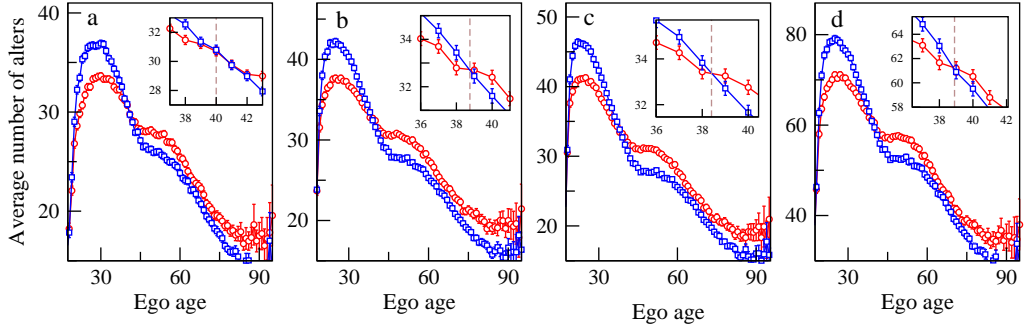}
\end{center}
\caption{The variation of the average number of alters with the age of the ego for different time windows. The ego networks in a, b and c are constructed from CDRs aggregated into the four month periods of January-April, May-August and September-December, respectively. The figure d corresponds to the time  window of the whole year. The figure legend is the same as that in Fig. \ref{fig-1}b. The figures in the inset focus on the region  where the crossover in behaviour for males and females is found. The dashed lines are used to denote the age of the crossover in each case.}
\label{fig-2}
\end{figure}

\section*{Methods}
\label{sec:data}
We analyze anonymized CDRs from a particular operator in a European country during 
2007. The CDRs contain full calling histories for  the subscribers of this operator (we term them `company users' and subscribers of other operators `non-company users'). There are $6.6$ million company users and around $25$ million non-company users appearing in the CDRs in the full one year period. Out of the total set of company users there are $3.2$ million users for whom both the age and the gender are available and only a single subscription is registered. In this study, we have only focused on the voice calls and excluded SMS entries from the CDRs. We construct an ego-alter pair if there is at least one call event between them during the observed time period. In general, we study calling patterns pertaining to pairs for whom age and gender are known. However, when the demographic information of the alters is not important for the analysis, we include individuals for whom this information is not known.

\noindent{\bf Additional filtering.} In the data set, there are company users for whom multiple subscriptions are found under the same contract numbers. For such users it is difficult to determine their real age and gender. We bypass this issue by considering the gender and age to be unknown for such users. The stored age of each company user corresponds to the year when the contract was signed; as the starting year of users' contracts ranged from 1998 to 2007, we updated the age of each user according to the number of years between the beginning of the year when the contract was signed and the first day of 2007. For some users, their contract starting date is unknown, so we add the average age-correction in the population, which was $3$ years (rounded from the actual value of $3.2$).

\section*{Results}
\subsection*{Number of alters with the age of the ego}

First we show the variation in the average number of alters that egos contact in a month, as a function of ego's age. From Fig. \ref{fig-1}a, we find that the number of alters reaches a maximum at an age of around $25$. This is followed by a decrease till an age of around $45$.  From age $45$, the number of alters contacted stabilizes for about a decade. After $55$, there is again a steady decrease. In Fig. \ref{fig-1}b,  we partition these data by gender. From the plot, it is clear that the average number of alters for males is greater than that for females for ages below $39$. But from the age of $39$ onwards we observe that the number of alters for females is greater than that for males. To check the robustness of this finding, we use time windows of different length, as shown in Fig. \ref{fig-2}. We observe a consistent pattern and that there is a crossover age at around $39$, irrespective of the time window used.

\begin{figure}[t]
\begin{center}
\includegraphics[width=0.8\textwidth]{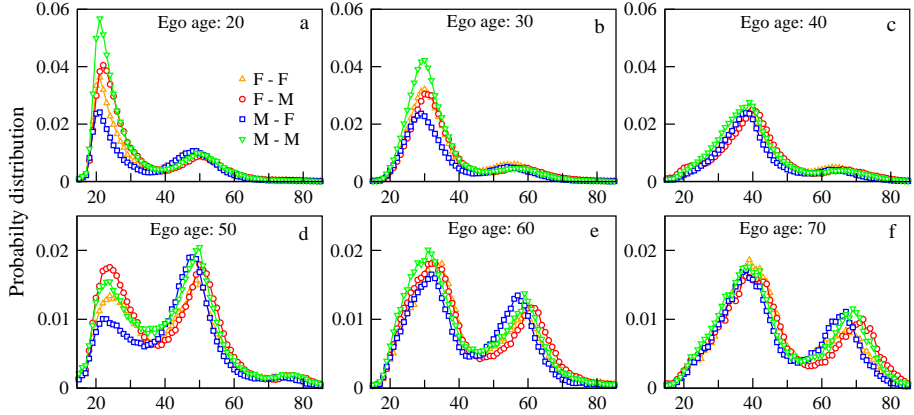}
\end{center}
\caption{Age distribution of alters for egos and alters of different sexes. The different plots correspond to the ego ages of 20 (a), 30 (b), 40 (c), 50 (d), 60 (e) and 70 (f). In each case we count the number of alters having different ages and normalize by the total number of alters (male and female). The different symbols and colours used to denote males (M) and females (F) in the ego-alter pairs, are orange triangle-ups (F--F), red circles (F--M), blue squares (M--F) and green triangle-downs (M--M). The distributions were calculated over a monthly time window and averaged over $12$ months. Only those CDRs were used where the demographic information of the egos as well as the alters was available.} 
\label{fig-3}
\end{figure}

\begin{figure}[h]
\begin{center}
\includegraphics[width=0.8\textwidth]{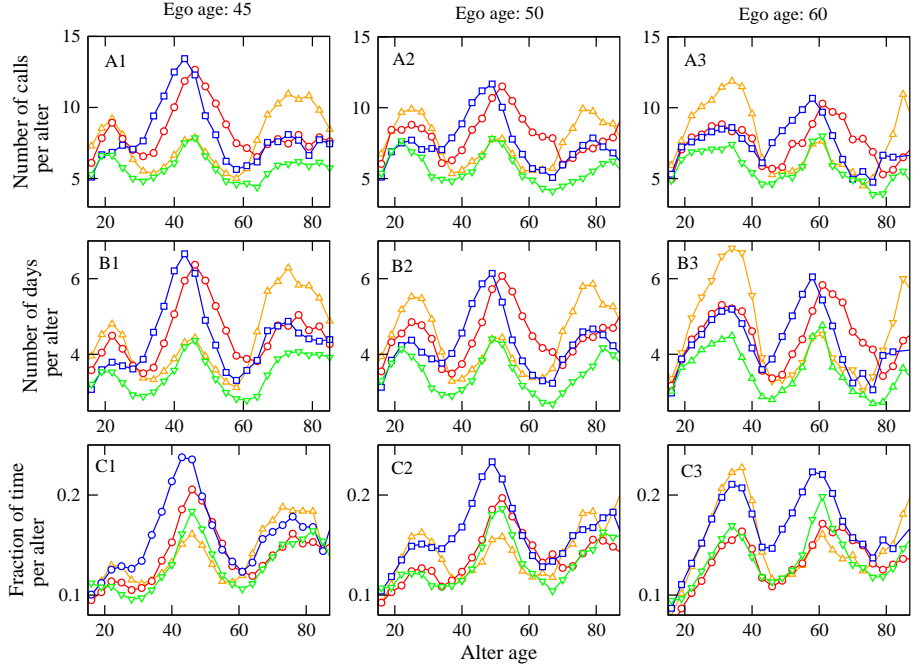}
\end{center}
\caption{Variation of different quantities characterizing the strength of communication, as a function of the age of alters for ego ages $45$ (A1, B1, C1), $50$ (A2, B2, C2) and $60$ (A3, B3, C3). The different symbols and colours denote the sexes in the ego-alter pairs and are similar to that used in Fig. \ref{fig-3}. The quantities are obtained from monthly call patterns and are averaged over the $12$ months period.}
\label{fig-4}
\end{figure}

\begin{figure}[h]
\begin{center}
\includegraphics[width=0.8\textwidth]{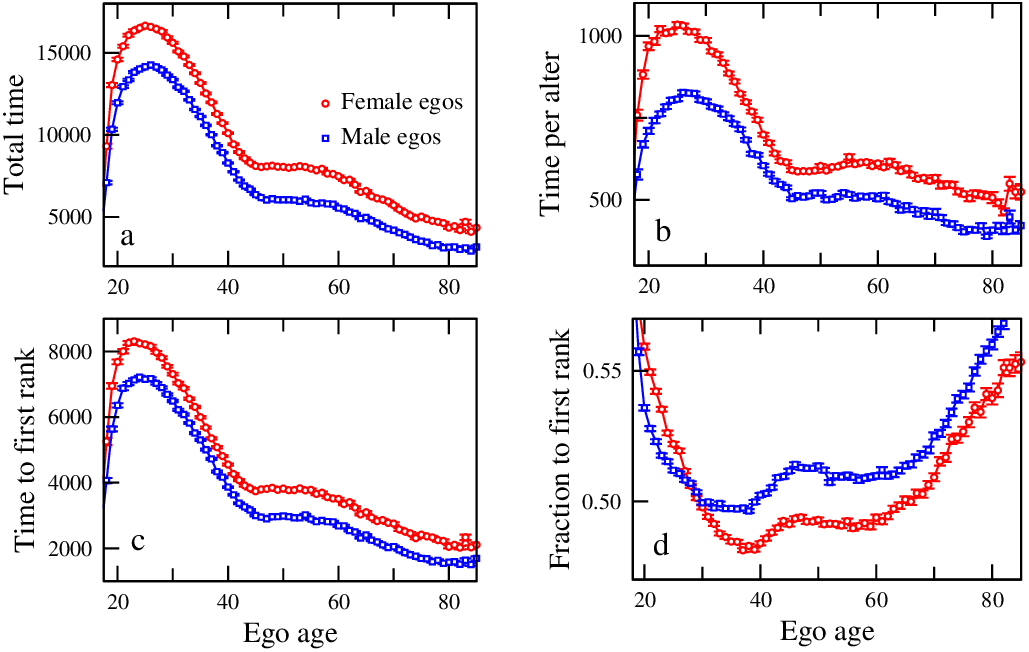}
\end{center}
\caption{Variation of different quantities characterizing the time budget of egos as a function of their age, for networks constructed in the time window of a month and averaged over the $12$ months: (a) total time (sec) per ego for all calls aggregated in the period, (b) time spent (sec) per alter per ego (sec), (c) time spent (sec) per ego with the first ranked alter, and (d) the fraction of the total time per ego that is spent with the first rank. Red circles and blue squares indicate female and male egos, respectively.}
\label{fig-5}
\end{figure}

\subsection*{Interaction probability and strength of interaction}

To investigate the interaction pattern of the egos belonging to different age groups, we measure the probability of interaction as a function of the age of alters. For egos of a given age, we find this probability by calculating the number of alters of any age and sex and divide by the total number of alters (male and female). In Fig. \ref{fig-3} we plot the distribution for egos belonging to six different decadal age classes, namely $20$s, $30$s, $40$s, $50$s, $60$s and $70$s. In general, the distributions appear to be have double peaks. The difference between the ages at which the peaks appear is around $25$ years. This is roughly a generation gap and is similar to the results in \cite{palchykov2012sex} where the age distribution of the most frequently contacted alter was investigated. Notice that the focus of the peaks differs with increasing age: in the younger age groups, the main peak is on individuals of the same age (peers), but from age $50$ this starts to be replaced by an increasingly large peak that is a generation younger than ego (presumably ego's now adult offspring). Note how these peaks track each other across the age space as ego ages. Note also the asymmetry in calling pattern between parent and child: $50$-year-olds (the parents) call $25$-year-olds (their adult children) more than twice as often as the $25$-year-olds call them.

This probability distribution is based on counting the number of alters at any given age. To examine in detail the appearance of the third peak, we quantify the strength of the interaction between the egos of age around $50$ and their alters at different ages. We consider egos of age $45$, $50$ and $60$ years and measure the following quantities in the time window of a month: (i) number of calls per alter, (ii) number of distinct days each alter is contacted, and (iii) calling time per alter (time of all calls aggregated within monthly window). However, the total calling time fluctuates strongly, so in lieu of (iii), we express the monthly aggregated duration of phone calls to an alter for a given ego as a fraction of the total calling time of that ego. In Fig. \ref{fig-4} we show these three quantities as a function of the age of the alters, averaged over $12$ months. The plot shows the conspicuous presence of three peaks of comparable heights. For older alters (those aged $70$ years or more) the averages are inevitably affected by the small number of older mobile phone users. Nonetheless, in general, the strength of communication appears to be larger when the alter is of the opposite gender and of similar age (compare the plots corresponding to female-ego-to-male-alters [red circles] and male-ego-to-female-alters [blue squares]).  
 
\subsection*{Time budgets of males and females}

The structure of the communication pattern of egos is also reflected in the variation in the  monthly aggregated call durations. In Fig. \ref{fig-5}a and \ref{fig-5}b we plot the total calling time of egos and the calling time per alter, respectively. The plots show that females have larger total calling time as well as larger time per alter than males. Interestingly, the crossover in the number of alters (Fig. \ref{fig-1}b) does not get translated to the calling time per alter. For a given ego, we rank the alters in terms of the monthly calling times and plot the calling time to the first rank alter (Fig. \ref{fig-5}c). The time spent by an ego with the first rank is approximately $7$ times the time spent with an average alter. However, the variation in these quantities is very similar to that of the dependence of number of alters on the age of the ego. When the calling time to the first rank is expressed as a fraction of the total calling time, we observe three broad regimes, a rapid decrease till $40$ years of age, a slow variation in the range $40$--$60$ years and a steady rise from $60$ years onwards. Note that the variation over the whole age range is only $10\%$ of the average value which is around $0.5$. Additionally, a crossover in the behaviour of males and females is visible at the age of $27$.

\begin{figure}[t]
\begin{center}
\includegraphics[width=0.8\textwidth]{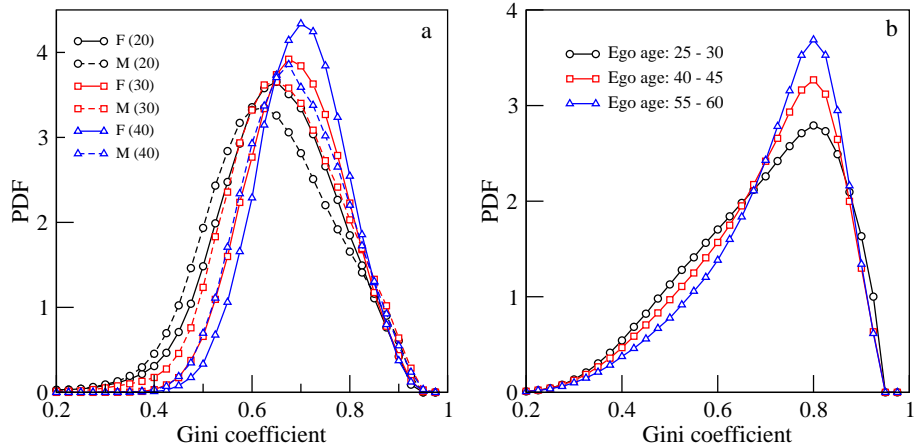}
\end{center}
\caption{The probability distribution function (PDF) of the Gini coefficient for different sexes and ages of egos. a. Distributions correspond to egos of different sexes (M: male, F: female) and having different number of alters (indicated inside brackets). b. Distributions corresponding to different age categories as indicated in the legend. Each distribution is calculated over a month and then averaged over $12$ months.} 

\label{fig-6}
\end{figure}

\subsection*{Inequality in call time distribution among alters}

Having discussed the dependence of mobile communication upon the gender and age of the egos, we provide a different perspective by measuring the inequality in the way social effort (indexed as calling times) is partitioned among the alters through the Gini coefficient for each ego. The Gini coefficient is mainly used to quantify income inequality and its value varies from $0$ (implying perfect equality) to $1$ (implying extreme inequality) \cite{coulter1989measuring}. We here use it as an evenness index: a Gini value of $0$ implies that ego devotes equal amounts of time with all the alters and a value of $1$ implies that the ego spends all the time with only one alter. 

It is natural to expect that there should be a strong bias among the egos with regard to the calling time spent with the alters, during a certain period of time. Here we analyse the nature of this bias by using the Gini coefficient in two different ways. We note that, for egos of a given age, there is a typical value for the number of alters. This poses a difficulty in comparing two egos of different ages because the value of the coefficient is known to depend upon the size of the sample set. We circumvent this issue in the following way. First, we consider egos irrespective of their ages but having a fixed number of alters, and calculate the Gini coefficient for the set of their monthly aggregated call times to the alters. Fig. \ref{fig-6}a plots the distributions for sets of egos having different genders. Comparison between the locations of the peaks suggest that females have overall higher Gini values compared to males.

Next, we consider egos irrespective of their gender. We choose egos in the following age brackets: (i) $25$--$30$, (ii) $40$--$45$ and (iii) $55$--$60$. For each ego we rank the alters with respect to the time of monthly aggregated call durations. Then we choose the call times belonging to the top $20$  alters. For egos having less than $20$ alters we assume the missing call times to be zero. However, in the analysis we exclude all egos who have less than $6$ alters. The resulting distribution of Gini values is shown in Fig. \ref{fig-6}b. We observe that the inequality among alters is larger for older people than for younger ones. The distributions in Fig. \ref{fig-6}b suggest that social effort becomes progressively less evenly distributed as people get older, and that this is true for both genders. In other words, older people devote more attention to their first ranked alters than younger people do. In effect, younger people are socially more promiscuous, but as they age they focus more and more of their effort, or social capital, on a smaller subset of meaningful relationships. As it is likely that most of an ego's first few rank alters are family members, this might suggest that older people become more attached to their family compared to younger people. Overall, the female egos exhibit higher inequality values than males do, and this suggests that females may be not only more socially focused than males, but also more attached to their family (as folk wisdom would also suggest).

\section*{Discussion}
\label{sec:summary}
\noindent

In order to explore the patterns of social investment across the lifespan in humans, we studied the records of mobile communication belonging to a particular European operator over a one year period. As these records include information on the service subscribers' age and gender, we are able to elucidate the nature of the interactions across the lifecycle. One important conclusion we can draw is that the average number of contacts is quite modest: in most cases, people focus their (phone-based) social effort each month on around $15$ people. This corresponds rather closely to the size of the second layer of egocentric personal networks in the face-to-face world \cite{zhou2005discrete,sutcliffe2012relationships}. In the face-to-face world, this layer also represents the number of alters contacted at least once a month. Thus, we provide some evidence that the use of mobile phone technology does not change our social world. It also provides further indirect evidence for the fact that we use the phone to contact those who are emotionally closest to us rather than simply those who live furthest away (see also \cite{jo2014spatial}).      

Our main finding, however, is the fact that the maximum number of connections for both males and females occurs at the 
age of around $25$ (Fig. \ref{fig-1}). During this early phase, males appear to be more connected than females. After this, the number of alters decreases steadily for both 
genders, although the decrease is faster for males than for females. The different rates of decrease result in a crossover around the age of $39$ such that after $39$ females become more connected than males. Note, however, in the age group $45$--$55$, the number of alters stabilizes to a very conspicuous plateau for both males and females. Projecting the slopes for the two graphs before the plateau suggests that the plateau represents a `saving' of around two alters who are retained as monthly alters rather than being lost to the next layer of less frequent contact. The difference between the plateau heights for females and males is around $1.5$ alters when the time window corresponds to one month. This difference grows to $3$ and $5$ alters when the window size is increased to four and twelve months, respectively. Thus, there are two separate but interrelated phenomena: the plateau that appears in both sexes during this period and the difference between males and females in the number of alters contacted. Since this age cohort is that in which ego's children typically marry and begin to reproduce in their turn, one likely explanation for this plateau is that it reflects  the fact that parents are maintaining regular interaction with their adult children at a time when some of these might otherwise be lost. The difference between the sexes seems to be primarily due to the more frequent interactions by the females with their adult children and the children's spouses.  Also, females intimately interact with their own close family members (e.g. keeping grandparents up-dated on the children's activities) and the new in-laws created by their children's marital arrangements.

This shift in women's social focus once her offspring reach adulthood and start to reproduce themselves is suggested by the appearance of a rather clear secondary peak in the number of alters aged about a generation ($25$ years) younger that appears in the contacts of $50$-year-olds (Fig. \ref{fig-3}d). This is in contrast with the profiles of younger cohorts (those aged $20$--$40$ years) who show a small, but distinct, secondary peak about a generation older than themselves (presumably their own parents). The positions of the peaks in Fig. \ref{fig-4} tell us quite a lot about domestic arrangements. For example, for this same-age cohort the peaks in the F-M (circles) and M-F (squares) curves in Fig. \ref{fig-4} are slightly offset, with the M-F leading by about $3$ years. In other words, on average a woman's main same-age alter is three years older than she is, while that for a man is about three years younger. This is almost exactly the typical age difference between spouses in contemporary Europe, including the country from which our sample derives \cite{hancock2003changes,drefahl2010does}.  It seems likely that in Fig. \ref{fig-4} the peaks to the left  are ego's children and the peaks to the right are ego's own parents. This suggestion is reinforced by the fact that these peaks track each other across the age space as ego ages.  

In addition, we found another crossover when we looked at the fraction of the total calling time devoted to the top ranked alter. This crossover occurs during the reproductively active period and its location roughly  corresponds to the maxima in Fig. \ref{fig-1}. Note, that before the crossover, the fraction for females, in Fig. \ref{fig-5}d, is larger than that for males, even though their maximum number of alters is actually lower. As the most frequently contacted alter is typically of the opposite sex \cite{palchykov2012sex}, we assume this to be the spouse. Because the time costs of reproduction in humans are very high (and may continue to be high for nearly two decades until the children reach marriageable age), we expect that females give priority to their spouses rather than other kinds of peers (siblings, cousins, friends) during this period when their time (and energy) budgets are under intense pressure. As a consequence, they maintain fewer relationships compared to males of the same age whose investment in their preferred alter seems to be much lower. A similar pattern of withdrawal from casual relationships so as to invest their increasingly limited available time in core relationships as time budgets are squeezed by the foraging demands of parental investment (in this case, lactation) has been noted in baboons \cite{altmann2001baboon,dunbar1988maternal}.

These results also seem to reflect female mate choice, with females persistently targeting their spouse in order to maintain investment in their chosen mate once they have made a choice (see also \cite{palchykov2012sex}). Note that, when examined over the whole age range, the fraction varies little and remains around $0.5$ (Fig. \ref{fig-5}d). This observation suggests that across the lifespan, the fractional allocation for the top ranked alter (the spouse) remains conserved even though the absolute time budget decreases (as can be seen from Fig. \ref{fig-5}a). This is reminiscent of the finding by \cite{saramaki2014persistence}, who reported, for a much smaller dataset, that the proportional distribution of social effort across all alters in an ego's network remains remarkably constant over time despite considerable change in network membership.

More generally, Fig. \ref{fig-6} suggest that there was a marked difference in the evenness with which the two genders distributed their social effort, as well as a progressive shift towards being less even with age. Females seemed to be generally more focused in their social arrangements than males, targetting more of their social effort onto fewer alters. This is reminiscent of the finding in \cite{david2015women} that women appear to have a small number of extremely close same-sex friendships, whereas males do not (they typically have a larger number of more casual same-sex friendships).  In addition, both genders exhibit the same tendency to shift from being more socially promiscuous (a more even Gini value) early in life to a more uneven (higher Gini value) in their $40$s. Since family dominate the inner layers of most people's social networks \cite{sutcliffe2012relationships}, this would suggest an increasing focus on family and close friendship relationships with age. This might reflect the fact that family relationships are more robust and resilient than friendships, as well as the fact that they are much more important as sources of lifelong support \cite{burton2014hamilton,curry2013altruism}. In contrast, the greater social promiscuity of younger individuals could be interpreted as a phase of social sampling in which individuals explore the range of opportunities (both for friendships and for reproductive partners) available to them before finally settling down with those considered optimal or most valuable. In this respect, the younger individuals may be viewed as `careful shoppers' \cite{tullock1971coal} who continue to check out the availability of options, only later concentrating their social effort on a select set of preferred alters. 

One implication of this is that turnover (or churn) in network membership might start to fall dramatically at a particular point in the life cycle marked by a shift from this more promiscuous phase to the more stable phase associated with a reduced social network. Fig. \ref{fig-1} suggests that the mean number of alters contacted falls from $15$--$20$ during this early phase to $8$--$15$ after age $40$. Figs. \ref{fig-1} and \ref{fig-5}d suggest that this switch in social focus may start to occur by the end of the third decade of life, and may thus coincide with the onset of reproduction. The average age of women at first birth in Europe for the currently reproducing generation is around $29$ \cite{eurostat} and would fit well with this prediction.


\section*{Acknowledgements}
KB, AG and DM acknowledges project COSDYN, Academy of Finland for financial support. DM also acknowledges CONACYT, Mexico for support. RD is supported by an ERC Advanced grant. We thank J\'anos Kert\'esz, Tamas David-Barrett and Hang-Hyun Jo for helpful  discussions.  

\section*{Author contributions statement}

KB, AG and DM carried out the analysis of the data. All the authors were involved in designing the project and the preparation of the manuscript. 

\section*{Additional information}
The authors declare no competing financial interests.


\begin{thebibliography}{99}

\bibitem{international2014world}
International~Telecommunication Union.
\newblock {\em The World in 2011: ICT Facts and Figures}.
\newblock ITU, 2014.

\bibitem{wang2011human}
Dashun Wang, Dino Pedreschi, Chaoming Song, Fosca Giannotti, and Albert-Laszlo
  Barabasi.
\newblock Human mobility, social ties, and link prediction.
\newblock In {\em Proceedings of the 17th ACM SIGKDD international conference
  on Knowledge discovery and data mining}, pages 1100--1108. ACM, 2011.

\bibitem{bagrow2011collective}
James~P Bagrow, Dashun Wang, and Albert-Laszlo Barabasi.
\newblock Collective response of human populations to large-scale emergencies.
\newblock {\em PloS one}, 6(3):e17680, 2011.

\bibitem{blondel2015survey}
Vincent~D Blondel, Adeline Decuyper, and Gautier Krings.
\newblock A survey of results on mobile phone datasets analysis.
\newblock {\em arXiv preprint arXiv:1502.03406}, 2015.

\bibitem{saramaki2015seconds}
Jari Saram{\"a}ki and Esteban Moro.
\newblock From seconds to months: an overview of multi-scale dynamics of mobile
  telephone calls.
\newblock {\em The European Physical Journal B}, 88(6):1--10, 2015.

\bibitem{miritello2013time}
Giovanna Miritello, Esteban Moro, Rub{\'e}n Lara, Roc{\'\i}o
  Mart{\'\i}nez-L{\'o}pez, John Belchamber, Sam~GB Roberts, and Robin~IM
  Dunbar.
\newblock Time as a limited resource: Communication strategy in mobile phone
  networks.
\newblock {\em Social Networks}, 35(1):89--95, 2013.

\bibitem{saramaki2014persistence}
Jari Saram{\"a}ki, EA~Leicht, Eduardo L{\'o}pez, Sam~GB Roberts, Felix
  Reed-Tsochas, and Robin~IM Dunbar.
\newblock Persistence of social signatures in human communication.
\newblock {\em Proceedings of the National Academy of Sciences},
  111(3):942--947, 2014.

\bibitem{roberts2011communication}
Sam~GB Roberts and Robin~IM Dunbar.
\newblock Communication in social networks: Effects of kinship, network size,
  and emotional closeness.
\newblock {\em Personal Relationships}, 18(3):439--452, 2011.

\bibitem{jo2014spatial}
Hang-Hyun Jo, Jari Saram{\"a}ki, Robin~IM Dunbar, and Kimmo Kaski.
\newblock Spatial patterns of close relationships across the lifespan.
\newblock {\em Scientific reports}, 4, 2014.

\bibitem{miritello2013limited}
Giovanna Miritello, Rub{\'e}n Lara, Manuel Cebrian, and Esteban Moro.
\newblock Limited communication capacity unveils strategies for human
  interaction.
\newblock {\em Scientific reports}, 3, 2013.

\bibitem{mcpherson2001birds}
Miller McPherson, Lynn Smith-Lovin, and James~M Cook.
\newblock Birds of a feather: Homophily in social networks.
\newblock {\em Annual review of sociology}, pages 415--444, 2001.

\bibitem{ip2006birds}
Grace Wai-man Ip, Chi-yue Chiu, and Ching Wan.
\newblock Birds of a feather and birds flocking together: physical versus
  behavioral cues may lead to trait-versus goal-based group perception.
\newblock {\em Journal of personality and social psychology}, 90(3):368, 2006.

\bibitem{palchykov2012sex}
Vasyl Palchykov, Kimmo Kaski, Janos Kert{\'e}sz, Albert-L{\'a}szl{\'o}
  Barab{\'a}si, and Robin~IM Dunbar.
\newblock Sex differences in intimate relationships.
\newblock {\em Scientific reports}, 2, 2012.

\bibitem{sutcliffe2012relationships}
Alistair Sutcliffe, Robin Dunbar, Jens Binder, and Holly Arrow.
\newblock Relationships and the social brain: integrating psychological and
  evolutionary perspectives.
\newblock {\em British journal of psychology}, 103(2):149--168, 2012.

\bibitem{dunbar2014reproductive}
Robin Dunbar.
\newblock {\em Reproductive decisions: an economic analysis of gelada baboon
  social strategies}.
\newblock Princeton University Press, 2014.

\bibitem{dunbar2010bondedness}
Robin~IM Dunbar and Susanne Shultz.
\newblock Bondedness and sociality.
\newblock {\em Behaviour}, 147(7):775--803, 2010.

\bibitem{lehmann2007group}
Julia Lehmann, AH~Korstjens, and RIM Dunbar.
\newblock Group size, grooming and social cohesion in primates.
\newblock {\em Animal Behaviour}, 74(6):1617--1629, 2007.

\bibitem{dunbar2009time}
RIM Dunbar, AH~Korstjens, and Julia Lehmann.
\newblock Time as an ecological constraint.
\newblock {\em Biological Reviews}, 84(3):413--429, 2009.

\bibitem{coulter1989measuring}
Philip~B Coulter.
\newblock {\em Measuring inequality: A methodological handbook}.
\newblock Westview Press, 1989.

\bibitem{zhou2005discrete}
W-X Zhou, Dider Sornette, Russell~A Hill, and Robin~IM Dunbar.
\newblock Discrete hierarchical organization of social group sizes.
\newblock {\em Proceedings of the Royal Society of London B: Biological
  Sciences}, 272(1561):439--444, 2005.

\bibitem{hancock2003changes}
Ruth Hancock, Rachel Stuchbury, and Cecilia Tomassini.
\newblock Changes in the distribution of marital age differences in england and
  wales, 1963 to 1998.
\newblock {\em POPULATION TRENDS-LONDON-}, pages 19--25, 2003.

\bibitem{drefahl2010does}
Sven Drefahl.
\newblock How does the age gap between partners affect their survival?
\newblock {\em Demography}, 47(2):313--326, 2010.

\bibitem{altmann2001baboon}
Jeanne Altmann.
\newblock {\em Baboon mothers and infants}.
\newblock University of Chicago Press, 2001.

\bibitem{dunbar1988maternal}
RIM Dunbar and Patsy Dunbar.
\newblock Maternal time budgets of gelada baboons.
\newblock {\em Animal Behaviour}, 36(4):970--980, 1988.

\bibitem{david2015women}
Tamas David-Barrett, Anna Rotkirch, James Carney, Isabel Behncke~Izquierdo,
  Jaimie~A Krems, Dylan Townley, and RI~Dunbar.
\newblock Women favour dyadic relationships, but men prefer clubs:
  Cross-cultural evidence from social networking.
\newblock {\em PloS one}, 10(3):e0118329, 2015.

\bibitem{burton2014hamilton}
Maxwell~N Burton-Chellew and Robin~IM Dunbar.
\newblock Hamilton’s rule predicts anticipated social support in humans.
\newblock {\em Behavioral Ecology}, page aru165, 2014.

\bibitem{curry2013altruism}
Oliver Curry, Sam~GB Roberts, and Robin~IM Dunbar.
\newblock Altruism in social networks: Evidence for a ‘kinship premium’.
\newblock {\em British Journal of Psychology}, 104(2):283--295, 2013.

\bibitem{tullock1971coal}
Gordon Tullock.
\newblock The coal tit as a careful shopper.
\newblock {\em American Naturalist}, pages 77--80, 1971.

\bibitem{eurostat}
Eurostat.
\newblock Population (demography, migration and projections).
\newblock
  http://ec.europa.eu/eurostat/web/population-demography-migration-projections/overview,
  2015.

\end{thebibliography}

\end{document}